\documentclass[letters]{mn2e}

\usepackage[fleqn]{amsmath}

\pagerange{L1--L5}

\newcommand{\beq}{\begin{equation}}
\newcommand{\eeq}{\end{equation}}
\newcommand{\bea}{\begin{eqnarray}}
\newcommand{\eea}{\end{eqnarray}}
\newcommand{\vc}[1]{{\textbf{#1}}}
\newcommand{\mc}[1]{\mathcal{#1}}
\newcommand{\cH}{\mathcal{H}}

\newcommand{\dd}{{\rm d}}
\newcommand{\fett}[1]{\boldsymbol{#1}}
\newcommand{\nabx}{\boldsymbol{\nabla}_{\boldsymbol{x}}}
\newcommand{\nabq}{\boldsymbol{\nabla}_{\boldsymbol{q}}}

\begin{document}

\title{Zel'dovich approximation and General Relativity}

\author[Cornelius Rampf and Gerasimos Rigopoulos]
{Cornelius Rampf$\,^a$\thanks{rampf@physik.rwth-aachen.de} 
and Gerasimos Rigopoulos$^{\,a,b}$\thanks{g.rigopoulos@tum.de} \\
$^a$ Institut f\"ur Theoretische Teilchenphysik und Kosmologie, Physikzentrum RWTH-Melaten, RWTH Aachen, D - 52056, Germany\\
$^b$ Physik Department T70,
James-Franck-Strasse 1, Technische Universit\"{a}t M\"{u}nchen,
D - 85748 Garching, Germany}

\maketitle

\begin{abstract}
We show how the Zel'dovich approximation and the second order displacement field of Lagrangian perturbation theory can be obtained from a
general relativistic gradient expansion in $\Lambda$CDM cosmology. The displacement field arises as a result of a second order non-local coordinate transformation which brings the synchronous/comoving metric into a Newtonian form.
We find that, with a small modification, the Zel'dovich approximation holds even on scales comparable to the horizon. The corresponding density perturbation is not related to the Newtonian potential via the usual Poisson equation but via a modified Helmholtz equation. This is a consequence of causality not present in the Newtonian theory. The second order displacement field receives relativistic corrections that are subdominant on short scales but are comparable to the second order Newtonian result on scales approaching the horizon. The corrections are easy to include when setting up initial conditions in large N-body simulations.
\end{abstract}

\begin{keywords}
cosmology: theory -- large scale structure of Universe -- dark matter
\end{keywords}

\section{Introduction}

The Zel'dovich approximation (ZA) \cite{Zeldovich:1969sb,Buchert:1987xy,Buchert:1992ya,Bouchet:1994xp,Buchert:1993xz,Bernardeau:2001qr,BuchertRampf:2012,Rampf:2012up}  provides a very simple analytical model of the gravitational evolution of CDM inhomogeneities which reproduces the appearance of the cosmic web, correlating well with the large scale filamentary features and void regions emerging in non-linear N-body simulations \cite{Sathyaprakash:1994wb,Buchert:1995km,Springel:2005mi,Matsubara:2007wj,Neyrinck:2012bf,RampfWong:2012} with the same initial conditions. The ZA can be derived from the full system of (Newtonian) gravitational equations and forms a subclass of solutions in Lagrangian perturbation theory (LPT) \cite{Buchert:1992ya}. In fact, the ZA and its second order improvement (2LPT) are  used to provide the initial displacements and velocities of particles in N-body simulations \cite{Scoccimarro:1997gr,Crocce:2006ve}.

The ZA arises in Newtonian theory and one might wonder about its status within general 
relativity \cite{Ellis:2002tq,Buchert:2012mb}. 
This question is particularly relevant if the ZA is used for example to set the initial dynamics of very large simulations which approach or exceed the size of the horizon. In this letter we show how the ZA and the 2LPT displacement field are derived in a general relativistic framework for $\Lambda$CDM cosmology. They correspond to a gradient expansion solution of the Einstein equations \cite{Rigopoulos:2012xj}, expressed in a coordinate system in which the metric takes a Newtonian form. In the process we calculate the relativistic corrections to the displacement field as well as the time shift between the proper time of the irrotational CDM particles and the ``Newtonian'' time corresponding to a weakly perturbed metric.

As is the case in LPT, the gradient expansion allows in principle for density contrasts that are larger than unity, $\delta{\rho} / \bar{\rho}>1$. The expansion eventually breaks down at points where caustics occur and the density becomes infinite. Close to such singularities higher order terms in the gradient series become important and the expansion loses its predictive power. However, one would expect that, unless black holes form, such singularities are in some sense removable since they appear where the worldlines of CDM particles cross. We find that when the gradient expansion breaks down, the corresponding Newtonian frame spacetime can still be considered a weak perturbation of FLRW.

\section{The gradient expansion metric}

The gradient expansion is a technique for approximating solutions to the Einstein equations which is not based on expanding in small perturbations, as in conventional perturbation theory, but on writing the time-evolved metric in terms of a series in powers of the initial 3-curvature. The idea 
dates back to 
Lifshitz \& Khalatnikov (1963) and Tomita (1975),
and was developed more recently in 
Stewart et al.~(1994), Comer et al.~(1994) and Tanaka \& Sasaki (2006) 
(see also Comer~(1996) and Bruni \& Sopuerta (2003) 
for covariant formulations). We will use here the gradient expansion solution for an irrotational flow of CDM particles in the presence of $\Lambda$ \cite{Rigopoulos:2012xj} to derive the Zel'dovich approximation.\footnote{See Barrow \& Goetz (1989) 
for a Newtonian treatment.}

Let us begin by writing the metric in synchronous comoving coordinates, possible to construct in this case,
\begin{align} 
 \label{co-synch}
 &\dd s^2 =-\dd t^2+\gamma_{ij}(t,\vc{q}) \, \dd q^{i} \dd q^{j}\,.
\end{align}
Summation over repeated spatial indices is implied.
Here $t$ is the proper time of the CDM particles and $\vc{q}$ are comoving coordinates, constant for each CDM fluid element. The metric can then be approximated by \cite{Rigopoulos:2012xj}
\begin{align}
&\gamma_{ij}(a,\vc{q}) \simeq a^2k_{ij} + \lambda(a) \left[ \hat{R}k_{ij}-4\hat{R}_{ij} \right]\nonumber\\
&+\!a^2\int\limits^a_0 \!\dd x \,\frac{\lambda(x) J(x) }{x^5H(x)}
  \,\Big[ 8\hat{R}^2k_{ij}-12\hat{R}^{kl}\hat{R}_{kl}k_{ij} \Big. \nonumber \\ 
     &\qquad \hspace{3.6cm} \Big. -28\hat{R}\hat{R}_{ij}+48\hat{R}_{ik}\hat{R}^k{}_j \Big]\nonumber\\
&-\!a^2\int\limits^a_0 \!\dd x \, \frac{L(x)}{x^5H(x)}
  \,\Bigg[ \frac{23}{4}\hat{R}^2k_{ij}-10\hat{R}^{kl}\hat{R}_{kl}k_{ij} \Bigg. \nonumber \\
  &\qquad \hspace{3.6cm} \Bigg. -18\hat{R}\hat{R}_{ij}+32\hat{R}_{ik}\hat{R}^k{}_j \Bigg] \nonumber\\
&+\!a^2\!\!\int\limits^a_0 \!\dd x \, \frac{\lambda(x) J(x)-L(x) }{x^5H(x)}
  \, 2 \!\left[ \hat{R}^{;k}{}_{;k}k_{ij}\!-\!4\hat{R}_{ij}{}^{;k}{}_{;k}\!+\!\hat{R}_{;ij} \right]\!.\! \label{eq:4thordersolution} 
\end{align}
In this expression $k_{ij}$ is an initial ``seed'' conformal metric describing the geometry early in the matter era. We assume this initial conditions to hold as $a\rightarrow 0$, effectively setting the lower limit of the integrals in (\ref{eq:4thordersolution}) to zero. Terms containing decaying modes have been neglected. Hats indicate that the curvature tensors are to be evaluated
from the initial time-independent conformal metric $k_{ij}$, e.g.~$\hat{R}=k^{ij}R_{ij}(k_{kl})$, and
a semicolon ``$;k$'' denotes a covariant derivative w.r.t.~this metric.
We have used the background FLRW scale factor $a(t)$ as the time variable and
\begin{align}
  H(a) &=H_0\sqrt{\Omega_M a^{-3}+\Omega_\Lambda} \,.
\end{align}
In terms of the proper time $t$ of the CDM particles we have 
\begin{align}
 a(t)=\exp\left\{ \int_{0}^t \dd t' \,H(t')\right\}\,,
\end{align}
with
\begin{align}
 H(t)&=\sqrt{\frac{\Lambda}{3}}\coth\left(\frac{\sqrt{3\Lambda}}{2}\,\,t\right)\,.
\end{align}
The functions appearing in the integrands in~(\ref{eq:4thordersolution}) satisfy
\bea
\frac{\dd J}{\dd a}+ \frac{J}{a}=\frac{1}{2aH}\,,\,\,\,\,
\frac{\dd \lambda}{\dd a}-2\frac{\lambda}{a} =\frac{J}{aH}\,,\,\,\,\,
\frac{\dd L}{\dd a}- \frac{L}{a}=\frac{J^2}{aH} \,.\!\!
\eea
It is easy to write down the solutions to these as integral expressions but it is simpler numerically to solve the above equations directly. At early times, when the contribution from $\Lambda$ is negligible, we have
\bea\label{SCDM}
J\simeq\frac{a^{3/2}}{5H_0\sqrt{\Omega_M}}\,,\quad \lambda\simeq\frac{a^3}{5H_0^2\Omega_M}\,,\quad L\simeq\frac{2}{175}\frac{a^{9/2}}{H_0^3\Omega_M^{3/2}} \,,
\eea
and $a\simeq t^{2/3}\left(H_0\sqrt{\Omega_M}\right)^{2/3}$. These expressions are exact for an EdS universe with $\Omega_M \rightarrow 1$ and $H_0 \rightarrow 2/(3t_0)$.

Let us now focus on the following conformal seed metric
\beq\label{initial-metric}
k_{ij}=\delta_{ij}\left[ 1+\frac{10}{3}\Phi(\vc{q}) \right] \,,
\eeq
where $\Phi(\vc{q})$ is the initial Newtonian potential, taken to be a Gaussian random field with amplitude given by the appropriate transfer function. The metric $k_{ij}$ is simply the linear initial condition derived from inflation and expressed in the synchronous gauge. Of course,
 non-Gaussian initial conditions could also be incorporated in $k_{ij}$ but we keep (\ref{initial-metric}) for simplicity. Solution (\ref{eq:4thordersolution}) becomes
\begin{align}
\gamma_{ij} &\simeq a^2\delta_{ij}\left[ 1+\frac{10}{3}\Phi(\vc{q})\right] \nonumber \\
  + & \,\,\frac{20}{3} \lambda(a)\left[\Phi_{,ij}\left(1-\frac{10}{3}\Phi\right)
 -5\Phi_{,i}\Phi_{,j}+\frac{5}{6}\delta_{ij}\Phi_{,l}\Phi_{,l} \right]\nonumber\\
&\quad\,\,\,\,+\,T_1(a) \,\Phi_{,li}\Phi_{,lj} -T_2(a)\, \Phi_{,ll}\Phi_{,ij}- \frac{T_2(a)}{4} F  \delta_{ij} \label{metric-phi}\,,
\end{align}
where
\begin{align}
 T_1 &=a^2\int\limits^a_0 \frac{\dd x}{x^5H(x)}\frac{200}{3} \left[\lambda(x) J(x)-\frac{1}{3}L(x)\right] \,, \\
 T_2 &=a^2\int\limits^a_0 \frac{\dd x}{x^5H(x)}\frac{400}{9}\left[\lambda(x) J(x)-\frac{1}{2}L(x)\right] \,,
\end{align}
and
\begin{align}
 F&= \Phi_{,lm}\Phi_{,lm}-\Phi_{,ll}\Phi_{,mm} \,.
\end{align}
A ``$,l$'' denotes a differentiation w.r.t.~Lagrangian coordinate $q_l$.
Note that a similar solution in the context of second order perturbation theory for Einstein-De Sitter cosmology was given in
Matarrese, Pantano \& Saez (1994).
However, the expressions are not identical to ours;
we have retained terms with two spatial gradients (the first two lines of (\ref{metric-phi})) which are up to two powers in the potential $\Phi$.
These terms are crucial for the coordinate transformation below.

The energy density is given by
\beq\label{density}
\rho(a,\vc{q})=\frac{3H_0^2\Omega_M}{8\pi G}\frac{\left[ 1+\frac{10}{3}\Phi(\vc{q}) \right]^{3/2}}{\sqrt{\det \left[\gamma_{ij}(a,\vc{q}) \right]}} \,,
\eeq
which matches to the linear perturbation theory density in synchronous gauge at sufficiently early times. It should be stressed that in deriving (\ref{eq:4thordersolution}) no assumption has been made about the magnitude of the density perturbation and values of $\delta\rho / \bar{\rho} >1$ are in principle allowed. Of course, the density is accurate only up to the gradient order kept in the expression for the metric. However, the metric in the form (\ref{eq:4thordersolution}) or (\ref{metric-phi}) predicts that eventually regions of zero volume will form where the density becomes infinite.\footnote{See Rigopoulos \& Valkenburg (2012) 
 for the spherical case and the corresponding Zel'dovich approximation formula for the density.} As we will see below, this in general corresponds to the formation of caustics in LPT.

\section{Newtonian coordinates and the displacement field}

The spatial coordinate system used above is comoving with the CDM fluid, i.e., each fluid element, or particle, is characterised
by a fixed $\vc{q}$ throughout the evolution. All information about inter-particle distances and clustering is encoded in the metric.
This however is not the most convenient way to visualise the situation, to relate to Newtonian intuition or, for example,
to compare with the output of an N-body simulation. Let us therefore define a coordinate transformation from the
comoving coordinates $(t,\vc{q})$ to another coordinate system $(\tau, \vc{x})$ where we require the metric to take the Newtonian form
\begin{align}
g_{00}(\tau,\vc{x})&=-\left[ 1+2A(\tau,\vc{x}) \right] \,, \\
g_{0i}(\tau,\vc{x})&=0 \,, \\
g_{ij}(\tau,\vc{x})&=\delta_{ij}\left[ 1-2B(\tau,\vc{x}) \right ] a^2(\tau) \,,
\end{align}
where $A\ll 1$ and $B\ll 1$, and where
\begin{align}
x^i(t,\vc{q})  &= q^{i}+\mc{F}^i(t,\vc{q})\label{x-xfn} \,, \\
\tau (t,\vc{q})&= t+\mc{L}(t,\vc{q})\,.
\end{align}
The metrics are related through
\begin{align}
\gamma_{ij} &=-\frac{\partial \tau}{\partial q^{i}}\frac{\partial \tau}{\partial q^{j}} &\hskip-0.5cm\left(1+2A\right)+\frac{\partial x^{l}}{\partial q^{i}}\frac{\partial x^{m}}{\partial q^{j}}\delta_{lm}\left(1-2B\right)a^2 \,,\label{glm}\\
0&=-\frac{\partial \tau}{\partial t}\frac{\partial \tau}{\partial q^{i}}&\hskip-0.5cm\left(1+2A\right)+\frac{\partial x^{l}}{\partial t}\frac{\partial x^{m}}{\partial q^{i}}\delta_{lm}\left(1-2B\right)a^2\,,\label{g0i}\\
-1 &=-\frac{\partial \tau}{\partial t}\frac{\partial \tau}{\partial t}&\hskip-0.5cm\left(1+2A\right)+\frac{\partial x^{l}}{\partial t}\frac{\partial x^{m}}{\partial t}\delta_{lm}\left(1-2B\right)a^2 \,.\label{g00}
\end{align}
In the above equations the various functions are evaluated \emph{at the same spacetime point}, which we choose at this stage to label with the $(t,\vc{q})$ coordinates that parametrize the worldlines of the CDM particles. For simplicity we will ignore possible vector and tensor modes that are generated at next to leading order -- this can be straightforwardly rectified.
We can now obtain the displacement field and the time shift at different orders in the potentials
\begin{align}
\mc{F}^i&=\mc{F}^i_{1}(t,\vc{q})+\mc{F}^i_{2}(t,\vc{q})+\ldots\,, \\
\mc{L} &=\mc{L}_{1}(t,\vc{q})+\mc{L}_{2}(t,\vc{q})+\ldots\,.
\end{align}

\subsection{Zel'dovich approximation}

Solving (\ref{glm}) - (\ref{g00}) at linear order in $\Phi$ we obtain
\begin{equation}
 \label{GT1}
\mc{F}^i_1(t,\fett{q}) = \frac{10}{3}\frac{\lambda(t)}{a^2(t)}\frac{\partial}{\partial q^i}\Phi(\vc{q})
  \,,\,\,\, \mc{L}_1(t,\fett{q}) = \frac{10}{3}J(t)\, \Phi(\vc{q}) \,, \!\!
\eeq
and the gravitational potentials read
\beq \label{A1}
 A_1(\tau,\vc{x}) =B_1(\tau,\vc{x})=\frac{5}{3}\big[ 2H(\tau)J(\tau)-1\big]\Phi(\vc{x}) \,.
\eeq
We see that the transformation (\ref{x-xfn}) has a direct interpretation: When expressed in terms of $\tau$ it is simply the trajectory in the Newtonian frame $(\tau,\vc{x})$ of a particle with initial coordinate $\vc{q}$:
 \beq\label{Zeldo1}
 \vc{x}(\tau, \vc{q}) \simeq   \vc{q} + \frac{10}{3}\frac{\lambda(\tau)}{a^2(\tau)}\frac{\partial}{\partial\vc{q}}\Phi(\vc{q})\,,
 \eeq
where the replacement $t \rightarrow \tau$ only induces a change at second order. We have checked that the prefactor $10 \lambda /(3a^2)$,
although satisfying apparently different equations is numerically identical to the $\Lambda$CDM growth factor $D_+(\tau)$
\begin{align}
\frac{10}{3}\frac{\lambda(\tau)}{a^2(\tau)} &=D_+(\tau)\equiv\frac{5}{2}H_0^2\Omega_m\frac{\cH(a)}{a}\int\limits_{0}^a\frac{\dd x}{\cH^3(x)}
 \nonumber \\
&=\frac{2}{5\Omega_M^{3/2}}\,a^{5/2}\,
_2F_1\left(\frac{3}{2},\frac{5}{6};\frac{11}{6};-\frac{\Omega_\Lambda}{\Omega_m}a^3\right) \,,
\end{align}
with ${\cal H}$ the conformal Hubble parameter. The representation of $D_+$ in terms of the hypergeometric function $_2F_1$ was found in
Enqvist \& Rigopoulos (2011) and Belloso, Garcia-Bellido \& Sapone (2011).
We have thus obtained directly the Zel'dovich approximation for $\Lambda$CDM from a general relativistic solution. Note that the formal steps described above resemble a gauge transformation from the synchronous to the Newtonian gauge. However, we stress again that we have not assumed here that $\delta\rho / \overline\rho$ is smaller than unity. So, eq.~(\ref{Zeldo1}) applies also in principle when $\delta\rho /\overline\rho >1$.

Focusing on scales that are comparable to the Hubble length, we can expand (\ref{density}) to linear order in the potential and express the result in terms of $\tau$ using~(\ref{GT1}):
\beq\label{density-phi}
\frac{\delta\rho}{\overline\rho}(\tau,\vc{x})\simeq-\frac{\lambda(\tau)}{a^2(\tau)}\frac{10}{3}\fett{\nabla}_{\vc{x}}^2\Phi(\vc{x})+10 H(\tau) \, J(\tau) \, \Phi(\vc{x}) \,.
\eeq
We see that in the Newtonian frame and on scales comparable to the horizon the Newtonian potential and the density perturbation are related through a (modified) Helmholtz equation instead of the standard Poisson equation. Writing the equation in terms of an evolving Newtonian potential 
$\phi_{\rm N}(\tau,\vc{x})$ we have
\beq\label{helmholtz-newton}
\nabx^2\phi_{\rm N}(\tau,\vc{x})-3\frac{a^2HJ}{\lambda} \phi_{\rm N}(\tau,\vc{x})=\frac{\delta\rho}{\bar{\rho}} \,,
\eeq
with the non-local solution
\beq
\phi_{\rm N}(\tau,\vc{x})= - \!\int \dd^3\vc{y}\,\frac{\exp\left\{ -\sqrt{\frac{3a^2HJ}{\lambda}}|\vc{x}-\vc{y}| \right\}}{4\pi|\vc{x}-\vc{y}|}\frac{\delta\rho}{\bar{\rho}}(\tau,\vc{y}) \,. \!
\eeq
To be concrete let us set $\Omega_\Lambda \!=\!0$, $\Omega_{\rm m} \!=\!1$ so that eq.~(\ref{SCDM}) holds exactly. We then have
\beq
\frac{3a^2HJ}{\lambda}\rightarrow  \frac{3H_0^2}{a} \,.
\eeq
We see that density fluctuations at distances sufficiently far away do not contribute to the potential. Furthermore, the region over which
density fluctuations do contribute to the potential grows with time. This is of course expected since equation~(\ref{density-phi})
and the underlying solution~(\ref{Zeldo1}) are derived from general relativity. Such causal behaviour is absent in the Newtonian theory which misses the second term on the LHS of~(\ref{helmholtz-newton}).

Let us finally see how formula (\ref{density-phi}) can be understood in terms of the particle trajectories (\ref{Zeldo1}). Suppose for the moment that the Zel'dovich displacement (\ref{Zeldo1}) is imposed on a Euclidean grid. This would result in a density fluctuation
\beq \label{eucl}
  \left(\frac{\delta\rho}{\overline\rho}\right)_{\rm Euclidean}= -\frac{\lambda(\tau)}{a^2(\tau)}\frac{10}{3}\nabx^2\Phi(\vc{x}) \,.
\eeq
The true spatial geometry of the Newtonian frame is not Euclidean and the density contrast acquires an extra term $\left(10 HJ-5\right)\Phi$ due to the change of spatial volume associated with (\ref{A1}). Comparing with (\ref{density-phi}) we see that to obtain the correct density a condition on the initial Lagrangian positions of the particles must be imposed. Indeed, assuming particles initially displaced by $\vc{x}_{\rm ini} = \vc{q} + \vc{c}(\vc{q})$ with
\beq
\nabq\cdot\vc{c}=-5\Phi\,.
\eeq
and evolved with (\ref{Zeldo1}) will reproduce the correct density. This is in agreement with the result of Chisari \& Zaldarriaga (2011).
A detailed discussion about the initial conditions can be found in Buchert (2011), Buchert, Nayet \& Wiegand (2012) and Rampf \& Rigopoulos (2012). 

\subsection{Trajectory at second order and short-scale behaviour}

The displacement field, the time shift and the gravitational potentials can be calculated to second order as well.
Explicit expressions can be found in the appendix. The trajectory in the Newtonian frame $(\tau,\vc{x})$ of a particle with initial coordinate $\vc{q}$ reads at second order
\begin{align}
\label{x2}
\vc{x}(\tau,\fett{q}) &\simeq   \vc{q}  + \frac{10}{3}\frac{\lambda(\tau)}{a^2(\tau)}\frac{\partial}{\partial\vc{q}}\Phi(\vc{q}) +\frac{1}{8}\frac{T_2(\tau)}{a^2(\tau)}\frac{\partial}{\partial \vc{q}}\frac{1}{\nabq^2}F \nonumber\\
&+\frac{50}{9}\frac{\left[ \lambda(\tau)+J^2(\tau)\right]}{a^2(\tau)}\frac{\partial}{\partial \vc{q}}\frac{1}{\nabq^2}\left(\Phi_{,l}\Phi_{,l}-\frac{3}{2}\frac{1}{\nabq^2}F\right) \nonumber \\
 & -\frac{50}{9}\frac{\left[ 2\lambda(\tau)+J^2(\tau) \right]}{a^2(\tau)}\frac{\partial}{\partial \vc{q}}\Phi^2 \,,
\end{align}
where $1/\nabq^2$ denotes the inverse Laplacian.
The last term in the first line of (\ref{x2}) is precisely the result from Newtonian 2LPT \cite{BuchertRampf:2012,Rampf:2012up,RR}. On small scales
 the second line in (\ref{x2}) is completely negligible and we obtain the complete Newtonian result,
showing that Newtonian dynamics on short scales produce the correct evolution.
This is of course not surprising, since general relativity is constructed to reproduce Newtonian physics in the appropriate limit. However, on scales approaching the horizon the last two terms in (\ref{x2}) become comparable to the second order Newtonian terms, the ratio between the two scaling roughly as $\frac{\rm Relativistic}{\rm Newtonian}\sim\frac{H_0^2}{k^2}$. This shows that, unlike the first order result, at second order general relativistic effects do have an impact on the trajectories of particles on such scales. In particular, any deviation from the Zel'dovich approximation computed with Newtonian dynamics on scales approaching the horizon will introduce errors. However, since dynamics on such scales are accurately described by expression~(\ref{x2}), it is easy to include these corrections in an N-body simulation. We give more quantitative details in Rampf \& Rigopoulos (2012).

Finally, let us make a few comments for the regime where the gradient expansion breaks down. On short scales the second order potentials read (see appendix)
\begin{align}
A_2&(\tau,\vc{x})=B_2(\tau,\vc{x})\simeq \frac{25}{9}\frac{\lambda}{a^2}\Phi_{|l}\Phi_{|l} \nonumber \\
&+\frac{25}{9}\frac{1}{a^2}\left(2\lambda HJ-\lambda-J^2-HL\right)\frac{1}{\nabx^2} \overline F+\mc{O}(\Phi^2) \,,
\end{align}
where we have dropped terms that are not enhanced by spatial gradients; ``$|l$'' denotes differentiation w.r.t.~Eulerian coordinates $x_l$, and $\overline F$ is the analogue to $F$ in Eulerian space. It is interesting to examine what happens to the metric potentials when the gradient expansion solution breaks down. To simplify the expressions let us set $\Omega_\Lambda\!=\!0$, $\Omega_{\rm m} \!=\!1$ so that eq.~(\ref{SCDM}) holds exactly. 
Expression (\ref{metric-phi}) predicts that the components of the synchronous metric will go to zero for fluctuations with highest wave-number $k$ approximately at a time defined by
\beq
\frac{4}{3}\frac{k^2}{H_0^2}a\Phi \sim 1 \,.
\eeq
At this point the spatial volume element of the comoving synchronous hypersurfaces goes to zero and the density becomes infinite. In the Newtonian frame this signifies the crossing of the CDM particle worldlines and the formation of caustics (shell crossings). At these points the gradient expansion solution breaks down.

Ultimately, shell crossings are the result of the assumption of a single velocity for each fluid element. In reality such singularities will be smoothed out by the non-zero velocity dispersion of the CDM particles but the zero pressure gradient expansion solution used here will be inaccurate in these regions. However, some qualitative estimates can be made. At the time when these singularities form we approximately have
\beq\label{singularity-potential}
A_2\sim B_2 \sim \frac{5}{84}\Phi \,.
\eeq
We thus see that the second order correction to the metric is enhanced, formally becoming first order. This would signify that, even if zero pressure is still assumed, the complete series should be summed to obtain the correct spacetime metric. However, unless terms of successive orders become even more dominant, eq.~(\ref{singularity-potential}) shows no evidence that spacetime in these high density regions will be significantly different from FRW. This statement of course would be incorrect close to the formation of black holes which cannot be seen in this formalism. But this should not be the case for most such regions.

\section{Summary and Discussion}

We have shown that the application to our universe of the gradient expansion method for approximating solutions to the Einstein equations is the
relativistic equivalent of solving Lagrangian Perturbation Theory. At first order the relativistic displacement field coincides with the Zel'dovich approximation up to an extra initial displacement $\fett{c}(\fett{q})$ which has to be imposed on the initial positions of particles to reproduce the correct density. We have therefore found that even for scales close to (or larger than) the horizon, the Zel'dovich approximation is essentially correct as a description of particle motion. However, the relation between the resulting density contrast and the Newtonian potential is not the standard Poisson equation but a modified Helmholtz equation, reflecting the causality of the relativistic theory. Contrary to what happens at first order, the second order displacement field receives relativistic corrections that are as important as the corresponding Newtonian result on large scales.

One can draw two main conclusions from the above findings. The first is that the fully relativistic solution reproduces the Newtonian dynamics on short scales. This is of course not surprising. However, we believe this is the first time it is explicitly shown starting from a fully relativistic solution, making no assumptions on the magnitude of the density perturbation. Our results show no evidence that Newtonian cosmology is not a good description on short scales. Correspondingly we expect any backreaction to be small even when large density contrasts form. This finding should be compared with the backreaction estimated in the synchronous gauge \cite{Kolb:2005da,Enqvist:2011aa}.

The second conclusion is that on large enough scales relativistic effects start contaminating the second order Newtonian result with the relative importance of the relativistic terms scaling as $H_0 / k^2$. Such corrections will be relevant for simulations that encompass the horizon. Since the Zel'dovich term will dominate on such scales, the corrections will be rather small. However, any deviation from the Zel'dovich approximation computed purely through Newtonian dynamics will miss the relativistic corrections in (\ref{x2}). Formula (\ref{x2}) then provides a direct way to include relativistic effects on the trajectory of particles in large N-body simulations. We will return to this issue with a more quantitative treatment in
Rampf \& Rigopoulos (2012). 

\section{Acknowledgements}
\noindent GR is supported by the Gottfried Wilhelm Leibniz programme of the Deutsche Forschungsgemeinschaft (DFG). We would like to thank Shaun Hotchkiss for useful discussions.

\section*{Appendix}

\noindent In this appendix we summarise our findings of the second order transformation to the Newtonian frame. It reads
\begin{align}
 \label{S2}
\mc{F}_2^i &(t,\vc{q}) = \frac{1}{8}\frac{T_2}{a^2}\frac{\partial}{\partial q^i}\frac{1}{\nabq^2}F-\frac{100}{9}\frac{\lambda}{a^2}\frac{\partial}{\partial q^i}\Phi^2 \nonumber \\ 
&\quad +\frac{50}{9}\left(J^2+\lambda\right)\frac{1}{a^2}\frac{\partial}{\partial q^i}\frac{1}{\nabq^2}\left(\Phi_{,l}\Phi_{,l}-\frac{3}{2}\frac{1}{\nabq^2}F\right) \,, 
\end{align}
\begin{align}
\label{L2}
\mc{L}_2&(t,\vc{q}) =\frac{50}{9}\frac{1}{a^2}\left(\lambda J-\frac{1}{2}L\right)\frac{1}{\nabq^2}F + \frac{50}{9}\frac{\lambda J}{a^2}\Phi_{,l}\Phi_{,l} \nonumber \\
&\quad+\frac{50}{3}\left(J-2HJ^2\right)\left(\frac{2}{3}\frac{1}{\nabq^2}\Phi_{,l}\Phi_{,l}-\frac{1}{\left(\nabq^2\right)^2}F\right) \nonumber \\
&\quad-\frac{25}{9}\left(J+2HJ^2\right)\Phi^2 \,,
\end{align}
{while the second order potentials are}
\begin{align}
\label{A}
A_2&(\tau,\vc{x})= \frac{25}{9}\frac{\lambda}{a^2}\Phi_{|l}\Phi_{|l}
 +\frac{25}{9}\left(\frac{3}{2}-\frac{18}{5}HJ+\frac{2}{5}J^2\Lambda\right)\Phi^2 \nonumber\\
 &\quad  +\frac{25}{9}\frac{1}{a^2}\left(2\lambda HJ-\lambda-J^2-HL\right)\frac{1}{\nabx^2}\overline F \nonumber \\
&\quad-\frac{50}{3}\left(\frac{1}{2}-3HJ+7H^2J^2-J^2\Lambda\right) \nonumber \\
&\qquad \hspace{2cm} \times \left(\frac{2}{3}\frac{1}{\nabx^2}\Phi_{|l}\Phi_{|l}-\frac{1}{\left(\nabx^2\right)^2}\overline F\right) \,, 
\end{align}
\begin{align}
B_2&(\tau,\vc{x}) = \frac{25}{9}\frac{\lambda}{a^2}\Phi_{|l}\Phi_{|l} +\frac{25}{9}\left(\frac{2}{5}HJ-8H^2J^2+\frac{2}{5}J^2\Lambda\right)\Phi^2\nonumber\\ 
&\quad+ \frac{25}{9}\frac{1}{a^2}\left(2\lambda HJ-\lambda-J^2-HL\right)
 \frac{1}{\nabx^2} \overline F \nonumber \\ 
&\quad +\frac{50}{3}HJ\left(1-2HJ\right)
\left(\frac{2}{3}\frac{1}{\nabx^2}\Phi_{|l}\Phi_{|l}-\frac{1}{\left(\nabx^2\right)^2} \overline F \right) \,.
\label{B}
\end{align}
Note again that the dependences and derivatives in eqs.~(\ref{S2}) and~(\ref{L2}) are w.r.t.~Lagrangian coordinates,
while for eqs.~(\ref{A}) and~(\ref{B})  w.r.t.~Eulerian coordinates.


\begin{thebibliography}{}
\bibitem[Barrow \& Goetz 1989]{1989CQGra...6.1253B}
Barrow J.\,D.,  Goetz G., 1989,
 Class.\,Quant.\,Grav., 6, 1253 

\bibitem[Belloso  et al.~2011]{Belloso:2011ms} 
  Belloso A.\,B., Garcia-Bellido J., Sapone D., 2011,
  {JCAP}, {1110}, 010 


\bibitem[Bernardeau et al.~2002]{Bernardeau:2001qr}
 Bernardeau F., Colombi S., Gaztanaga E., Scoccimarro R., 2002, {Phys.\,Rept.}, {367}


\bibitem[Bouchet et al.~1995]{Bouchet:1994xp} Bouchet F.\,R., Colombi S., Hivon E., Juszkiewicz R.,
 1995, {A\&\!A}, {296}, 575

\bibitem[Bruni \& Sopuerta 2003]{Bruni:2003hm}
  Bruni M., Sopuerta C.\,F., 2003,
  Class.\,Quant.\,Grav., 20, 5275

\bibitem[Buchert 1992]{Buchert:1992ya}  Buchert T., 1992, MNRAS, 254, 729

\bibitem[Buchert 2011]{Buchert:2011yu} Buchert T., 2011, Class.\,Quant.\,Grav., 28, 164007

\bibitem[Buchert \& Ehlers 1993]{Buchert:1993xz} Buchert T., Ehlers J., 1993,
  {MNRAS}, {264}, 375 

\bibitem[Buchert et al.~1997]{Buchert:1995km} Buchert T., Karakatsanis G., 
  Klaffl R., Schiller P., 1997, {A\&\!A}, {318}, 1 

\bibitem[Buchert et al.~2012]{BuchertWiegand} Buchert T., Nayet C., Wiegand A., 2012, in preparation

\bibitem[Buchert \& Goetz 1987]{Buchert:1987xy} Buchert T., Goetz G., 1987, 
   J.\,Math.\,Phys., 28, 2714

\bibitem[Buchert \& Ostermann 2012]{Buchert:2012mb} Buchert T., Ostermann M., 2012, 
  {Phys.\,Rev.} D, {86}, 023520 

\bibitem[Chisari \& Zaldarriaga 2011]{Chisari:2011iq}
  Chisari N. E., Zaldarriaga M., 2011,
  {Phys.\,Rev. D}, 83, 123505

\bibitem[Comer 1997]{Comer:1996du} Comer G. L., 1997,
  {Class.\,Quant.\,Grav.}, 14, 407

\bibitem[Comer et al.~1994]{Comer:1994np} Comer G. L., Deruelle N., Langlois D., Parry J., 1994,
  {Phys.\,Rev.\,D},  49, 2759

\bibitem[Crocce, Pueblas \& Scoccimarro 2006]{Crocce:2006ve} Crocce M., Pueblas S., Scoccimarro R., 
  2006, {MNRAS}, {373}, 369 

\bibitem[Ellis 2002]{Ellis:2002tq} Ellis G.\,F.\,R., Tsagas C.\,G., 2002,
  Phys.\,Rev.\,D, 66, 124015

\bibitem[Enqvist et al.~2012]{Enqvist:2011aa}
  Enqvist K., Hotchkiss S., Rigopoulos R., 2012, 
  {JCAP}, 1203, 026

\bibitem[Enqvist \& Rigopoulos 2011]{Enqvist:2010ex}
  Enqvist K., Rigopoulos G., 2011,
  JCAP, {1103}, 005 

\bibitem[Kolb et al.~2006]{Kolb:2005da} Kolb E. W., Matarrese S., Riotto A., 2006,
  {New J.\,Phys.},  8, 322

\bibitem[Lifshitz \& Khalatnikov 1963]{Lifshitz:1963ps}
  Lifshitz E.\,M., Khalatnikov I.\,M., 1963,
  {Adv.\,Phys.}, 12, 185 

\bibitem[Matarrese et al.~1994]{Matarrese:1994wa} Matarrese S., Pantano O.,  Saez D., 1994,
  MNRAS,  271, 513 

\bibitem[Matsubara 2008]{Matsubara:2007wj} Matsubara T., 2008, Phys.\,Rev.\,D, {77}, 063530 

\bibitem[Neyrinck 2012]{Neyrinck:2012bf} Neyrinck M., 2012, preprint (arXiv:1204.1326)

\bibitem[Rampf 2012]{Rampf:2012up} Rampf C., 2012, JCAP, 1212, 004

\bibitem[Rampf \& Buchert 2012]{BuchertRampf:2012} Rampf C., Buchert T., 2012, {JCAP}, {1206},  021

\bibitem[Rampf \& Rigopoulos 2012]{RR} Rampf C., Rigopoulos G., 2012, in preparation

\bibitem[Rampf \& Wong 2012]{RampfWong:2012} Rampf C.,  Wong Y. Y. Y., 2012, {JCAP}, {1206}, 018

\bibitem[Rigopoulos \& Valkenburg 2012]{Rigopoulos:2012xj}
  Rigopoulos G., Valkenburg W., 2012, Phys.\,Rev.\,D, {86}, 043523 


\bibitem[Sathyaprakash et al.~1995]{Sathyaprakash:1994wb} Sathyaprakash B.\,S., Sahni V., 
  Munshi D., Pogosian D., Melott A.\,L., 1995, MNRAS, {275}, 463

\bibitem[Scoccimarro 1998]{Scoccimarro:1997gr} Scoccimarro R., 1998, {MNRAS}, {299}, 1097

\bibitem[Springel 2005]{Springel:2005mi} Springel V., 2005, {MNRAS}, {364}, 1105

\bibitem[Stewart et al.~1994]{Stewart:1994wq}
  Stewart J.\,M., Salopek D.\,S., Croudace K.\,M., 1994,
  {MNRAS}, 271, 1005 

\bibitem[Tanaka \& Sasaki 2006]{Tanaka:2006zp} Tanaka Y., Sasaki M., 2007,
  {Prog.\,Theor.\,Phys.},  117, 633

\bibitem[Tomita 1975]{Tomita:1975kj} Tomita K., 1975, Prog.\,Theor.\,Phys., 54, 730

\bibitem[Zel'dovich 1970]{Zeldovich:1969sb} Zeldovich Y.\,B., 1970,  {A\&\!A}, 5, 84

\end{thebibliography}
\end{document}